# About: "Float-stacked graphene–PMMA laminate"

Anirban Kundu[1], Won Kyung Seong[1*], S. Kamal Jalali[2], Nicola M. Pugno[2,3], and Rodney S. Ruoff[1,4,5,6*]

[1]*Center for Multidimensional Carbon Materials (CMCM), Institute for Basic Science (IBS), Ulsan 44919, Republic of Korea.*

[2]*Laboratory for Bioinspired, Bionic, Nano, Meta Materials & Mechanics, Department of Civil, Environmental and Mechanical Engineering, University of Trento, Via Mesiano, 77, 38123 Trento, Italy.*

[3]*School of Engineering and Material Science, Queen Mary University of London, Mile End Road, London E1 4NS, UK.*

[4]*Department of Chemistry, Ulsan National Institute of Science and Technology (UNIST), Ulsan 44919, Republic of Korea.*

[5]*School of Energy and Chemical Engineering, Ulsan National Institute of Science and Technology (UNIST), Ulsan 44919, Republic of Korea.*

[6]*Department of Materials Science and Engineering, Ulsan National Institute of Science and Technology (UNIST), Ulsan 44919, Republic of Korea.*

Email: one2rang@gmail.com, wks1130@ibs.re.kr,  ruofflab@gmail.com, rsruoff@ibs.re.kr

## Abstract

We report scientific and technical queries regarding the article reported by Kim *et al.*,[1] on the mechanical properties of graphene-poly(methyl methacrylate) (PMMA) composites. Our analysis finds that the current experimental data is insufficient to fully support the conclusions presented in the article. We suggest the reported enhancement in Young's modulus and strength of the graphene-PMMA laminates (GPL) samples are mainly due to the heat treatment of the polymer rather than the incorporation of graphene. The Raman spectroscopy data (as per our analysis) for the GPL samples indicates that large cracks and defects were introduced during the hot rolling process used to fabricate the graphene-PMMA composite. We believe that the queries will aid the audience in better understanding the mechanical response of graphene-PMMA composites.

Kim et al.,[1] described a "float-stacking fabrication" method to make "graphene- PMMA laminates (GPL)". The authors report (referred to as Ref.1 hereafter) that this architecture enabled enhancing stiffness and strength well beyond PMMA alone, reporting values of tensile strength ($\sigma \sim 141$ MPa) and Young's modulus ($E \sim 5.37$ GPa), with a stated graphene volume fraction of 0.19%. On April 15, 2024, Dr. Anirban Kundu, Dr. Won Kyung Seong, and Prof. Rodney S. Ruoff had an initial discussion regarding the work reported in Ref. 1 by Kim *et al.,* and as questions arose about the analyses and results reported in Ref. 1, we have decided to contribute this article here.

**Experimental Data:**

The thickness of each of the GPL samples (that is, all the 36 samples for GPL-0 to GPL-100) is reported as ~18 µm (17.07-19.89 µm), see supplementary Table 1 of Ref. 1. But the SEM micrographs (Fig. 3b,c) of Ref. 1 show (to us, per our analysis of the micrographs of these 3 samples) thicknesses of 3.66 ± 0.10 µm (S-GPL), 2.95 ± 0.13 µm (Tg-GPL), and 2.35 ± 0.13 µm (GPL): much different than 18 µm, as shown in Fig. 1a. The thickness values of S-GPL and $T_g$-GPL are not provided in Ref. 1. The authors said that the thickness of the stacked samples was measured from cross-sectional SEM images. Fig. 1c and supplementary Fig. 5 of Ref. 1 shows other SEM micrographs of the GPL samples. Our analysis of these SEM micrographs that are shown in supplementary Fig. 5 of Ref. 1, yields average thickness values of GPL-0 to GPL 100 as 18.78 ± 0.49, 22.26 ± 0.55, 22.89 ± 0.29, 23.03 ± 0.50, 20.89 ± 0.67, and 25.11 ± 0.94 µm, respectively; and 28.42 ± 0.90 µm from Fig. 1c. We provide the points selected for our measurements of thickness in our Fig. 1 and note that the authors of Ref. 1 did not indicate where thickness measurements were made. Thickness values measured in our analysis are compared in supplementary Table 1 with the thickness values reported in Ref. 1.

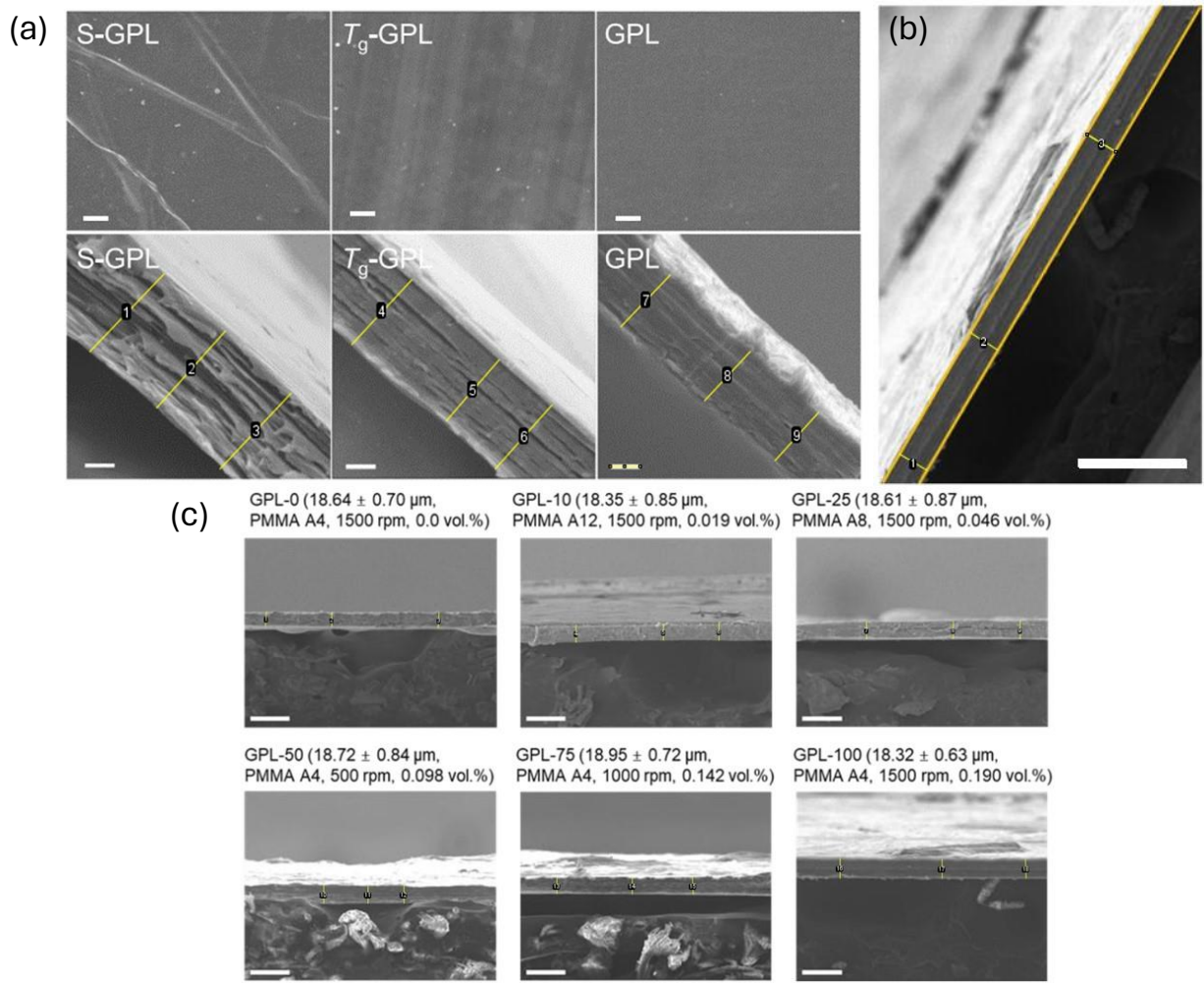

*Figure 1:* *Thickness measurements of the stacked graphene PMMA layer. (a) The thickness of S-GPL, $T_g$-GPL, and GPL is found to be 3.66 ± 0.10, 2.95 ± 0.13, and 2.35 ± 0.13 µm, respectively (image collected from Fig. 3c in Ref.1. (b) Thickness of GPL-100 measured from Fig. 1c. (c) Thickness measurement of GPL-0, GPL-10, GPL-25, GPL-50, GPL-75 and GPL-100 from supplementary Fig. 5. The thickness is measured along the straight line segments (yellow) marked in the figure.*

The stress-strain curves of three different samples (S-GPL, $T_g$-GPL, and GPL) with 25 graphene layers are shown in supplementary Fig. 9 of Ref. 1; we have analyzed these stress-strain curves (see Fig. 2) and find that our fitted $E$ values do not match with the values shown in supplementary Table 2 of Ref. 1. For GPL-25 the average $E$ was reported in supplementary Table 2 of Ref. 1 as 3.92 ± 0.22 GPa; we instead fit $E$ to be 6.59 ± 0.57 GPa. We have summarized the $E$ and $\sigma$ values obtained from our Fig. 2 (note that Fig. 2 is identical to supplementary Fig. 9 of Ref 1) in our supplementary Table 2. The actual $E$ for GPL-25 (obtained by us from supplementary Fig. 9) is thus higher than GPL-100, that Ref. 1 reports as the highest $E$ (5.37 ± 0.31 GPa). It is worth noting that the average failure strain for GPL-25 in Supplementary Fig. 9c and supplementary Table 3 of Ref. 1 is reported to be 1.90%, markedly differing from the value of 3.66% that one can readily fit from Fig. 2b of Ref. 1 (or that can be obtained from the excel file provided by Ref. 1).

Enhancement in $E$ and $\sigma$ values of GPL-100 is compared with GPL-0 in Ref. 1 to calculate the effective $E$ and $\sigma$ of the "graphene fillers" (as the graphene is referred to in Ref. 1) present in GPL. The enhancement (in GPL-100 compared to GPL-0) in tensile strength ($\sigma$) and Young's modulus ($E$) are reported in Ref. 1 as 277.5% (from 79.60 ± 4.10 to 141.29 ± 3.29 MPa) and 261.26% (3.33 ± 0.15 to 5.37 ± 0.31 GPa), respectively. The actual values are 77.5% (from 79.60 ± 4.10 to 141.29 ± 3.29 MPa) and 61.26 % (3.33 ± 0.15 to 5.37 ± 0.31 GPa), respectively, as per our analysis. The linear region considered for $E$ measurements in our analysis is shown with a red dotted line in Fig. 2.

The Raman mapping area is mentioned as 50 μm x 50 μm (Methods, Ref. 1) with a stated step size of 0.5 μm. Per the scalebar, the actual Raman mapping area (supplementary Fig. 13 in Ref.1) is 33 μm x 33 μm.

The graphene PMMA stacked layers were stated to be processed with a hot rolling press to prepare the GPL. The temperature of the hot rolling press is not mentioned in Ref. 1.

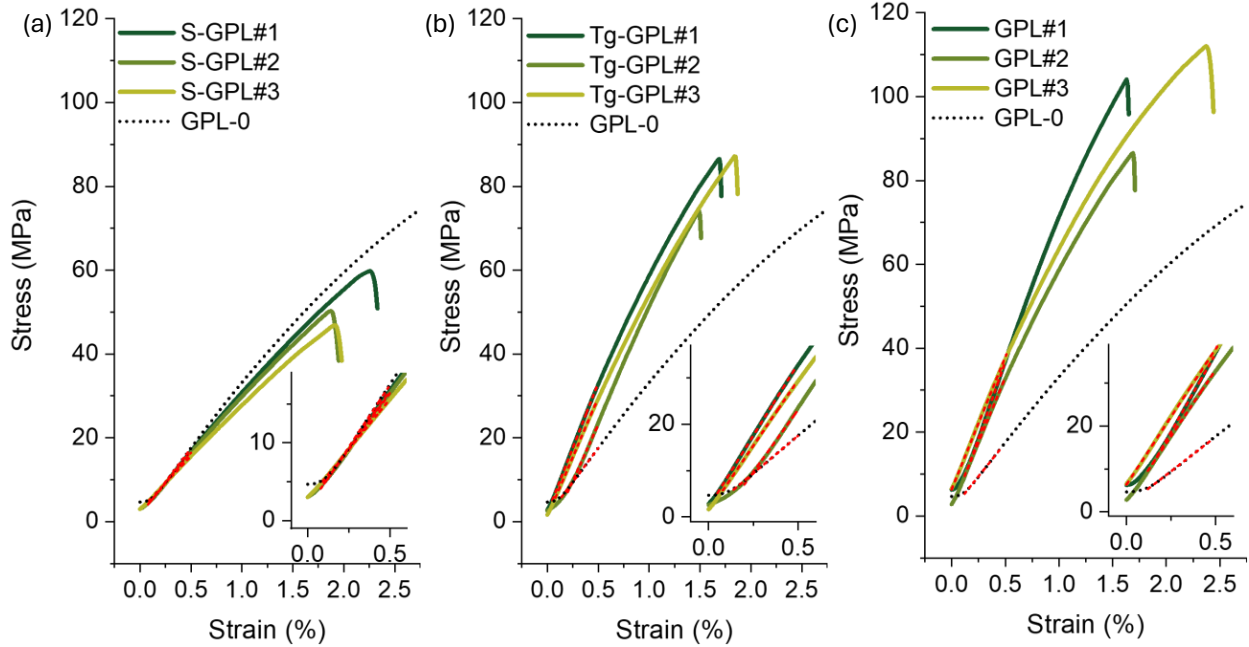

***Figure 2:*** *Analysis of (a) S-GPL, (b) Tg-GPL, and (c) GPL stress strain curves. The dotted curve represents the stress strain curve of GPL-0. The red dotted line depicts the linear fitted region to calculate the E. The inset plot shows the linear region of each curve below 0.5% applied strain.*

**Mechanical Properties Analysis:**

Ref. 1 reports Young's modulus ($E$) and strength ($\sigma$) values for graphene for sample GPL-100 of around 1.09 TPa and 33 GPa respectively, calculated using the rule of mixtures. However, it is not clear as per Ref. 1 whether these high values are for a single layer graphene[2,3], or of "graphene fillers" present in the PMMA matrix.

To clearly compare the mechanical properties of graphene-PMMA laminates (GPL) with pure PMMA (GPL-0), the GPL-0 samples need to be prepared with the same number of PMMA layers as the corresponding GPL samples, using the same float-stacking method reported in Ref. 1. The $E$ and $\sigma$ values of GPL-0 reported in Ref. 1 are close to those of bulk PMMA, suggesting that GPL-0 is made of PMMA that is like regular bulk PMMA. The thickness and stress-strain curves of stacked PMMA (S-GPL-0) and stacked PMMA above the glass transition temperature (T$_g$-GPL-0) are not reported in Ref. 1. However, since their thicknesses are comparable, we assumed that their mechanical properties are like that of GPL-0. We compared the stress-strain response of S-

GPL, T$_g$-GPL and GPL (all with 25 graphene layers) with GPL-0 in Fig. 2, from the data provided with Ref. 1. The $E$ and $\sigma$ values measured by us are summarized in supplementary Table 2. Our analysis based on Ref. 1 shows that the presence of graphene filler (25 monolayers) *has a negative effect* on the $E$ and $\sigma$ of S-GPL as compared to GPL-0. Whereas $\sigma$ is relatively decreased for T$_g$-GPL from GPL-0. GPL-25 shows an increase in both $E$ and $\sigma$ compared to GPL-0, as reported in Ref. 1, with the increase in $E$ being higher (6.59 GPa) than the reported enhancement in Ref. 1 for GPL-100 (5.37 GPa). Considering the stated volume fraction of 0.046% for graphene in GPL-25, such a substantial improvement in the Young's modulus of GPL-25, based on the rule of mixtures (Eq. 2 in Ref. 1), yields a calculated value of 7.35 TPa for the Young's modulus of graphene (but this is an impossibly high value). We discuss the effect of heat treatment on mechanical response of polymer in section 1 of Supplementary Information of this article. As reported by Zhang *et al.*[4], the mechanical properties of graphene are insensitive to the number of layers of graphene. However, the enhancement in $E$ and $\sigma$ with an increasing number of graphene layers stacked in GPL is reported in Ref. 1; this seems contradictory. Ref. 1 reports that as the number of graphene layers, $N$, increases while maintaining a constant volume fraction, the strength of GPL initially peaks and then diminishes with further layer increments. This pattern suggests the existence of an optimal number of layers to achieve the maximum strength. We have discussed this further in section 2 of Supplementary Information of this article.

**Structural Analysis:**

The reason(s) for the reported high mechanical properties of graphene in Ref. 1 is/are unclear, so it is important to consider the quality of graphene used in the graphene-PMMA laminates (GPL). The mechanical properties were stated to be measured on samples with a length of 10 mm and width of 3 mm per Ref. 1. However, Ref. 1 only shows one TEM image of a single graphene layer embedded in the PMMA matrix, which does not provide information about the overall graphene quality/morphology in GPL at different stages of sample preparation (from S-GPL to T$_g$-GPL and GPL). The SEM images (525 µm x 440 µm) showing the surface morphology of S-GPL, T$_g$-GPL, and GPL (in Fig. 3b of Ref. 1) also do not give information about the graphene quality (see Fig. 1a).

To understand the reinforcement effect of graphene in GPL, Raman mapping was performed on what was stated to be an 18-μm thick sample (the thickness value mentioned in Ref. 1). However, Raman mapping primarily probes the surface, and the depth (or volume) from which the spectra is collected depends on the instrument specifications. The Raman mapping shown over an area of 33 μm x 33 μm (supplementary Fig. 13 of Ref. 1) shows no evidence of grain boundaries. The observation of grain boundaries is expected for polycrystalline graphene in Raman mapping[5]. Raman depth profiling was reported in Ref. 1 (to evaluate the graphene quality). Our analysis of the depth profiling data (supplementary Fig. 3 in Ref. 1) suggests that it was done only at a single spot. The reported $I_{2D}/I_G$ ratio for the graphene transferred onto a silicon wafer is ~1.97 (calculated by us from supplementary Fig. 2 in Ref. 1), which is close to the value for monolayer graphene.[6] Usually, $I_{2D}/I_G > 2$ indicates good quality monolayer graphene, and lower values are for bilayer and multiple layers of graphene[6]. Our calculation shows that the depth profiling plot has an average $I_{2D}/I_G$ of 1.22 ± 0.07, indicating that the graphene in GPL is not in perfect monolayer condition and is 'mixed' (in some manner) with the PMMA matrix. Our analysis of the Raman depth profiling data (Supplementary Fig.3 of Ref. 1) found that the $I_{2D}/I_G$ value is also observed between the identified graphene layers (as per Ref. 1), and this value is comparable to pristine graphene as per Ref. 1. This further suggests that graphene is randomly distributed throughout the PMMA matrix, which is possible when the monolayer graphene is fragmented, perhaps due to smaller grain size and multiple grain boundaries. The hot rolling process (per Ref. 1) reportedly applies a force of 78.48 N (corresponding to a compressive pressure of 8.35 MPa) to remove voids, but this removal of trapped voids seems likely to introduce large cracks and defects in graphene. It appears that the monolayer graphene present in the polymer matrix was likely fractured and fragmented during the hot rolling process, forming a graphene-PMMA composite in which the graphene is actually 'flakes' (i.e., relatively small fragments) relative to a large area CVD grown graphene.

**Conclusion**

In conclusion, the float-stacking method proposed by Kim *et al.*[1] presents a promising approach for creating highly aligned graphene-polymer nanocomposites. However, there are several important issues (mentioned above) that need to be considered. It appears that the improvement in Young's modulus and strength of the stacked graphene-PMMA laminates (GPLs) reported in Ref.

1 is primarily or entirely the result of temperature-induced modification of PMMA, rather than the presence of graphene. Indeed, attributing the entire improvement to the latter leads to (we suggest) grossly overestimating the properties of the polycrystalline graphene.

## Author Contributions

All authors contributed to the analysis and interpretation of the literature that is discussed and jointly wrote the article.

## Corresponding Author

Correspondence to Won Kyung Seong and Rodney S. Ruoff.

## Acknowledgements

We appreciate Daniel Hedman and Sun Hwa Lee for reading and commenting on our manuscript.

# Supplementary Information File

**Section 1: Heating Effect on Polymer Mechanical Response**

Ref. 1 mentioned that the enhancement in $E$ and $\sigma$ in GPL is achieved by establishing good contact between graphene and PMMA by removing defects like wrinkles and voids through high temperature stacking (above the $T_g$ of PMMA) and hot rolling. The mechanical properties analysis of bare stacked PMMA samples (S-GPL-0 and Tg-GPL-0), are essential: but not presented in Ref. 1. Stacked PMMA samples prepared with varying layer numbers like the GPL ($N$=10, 25, 50, 75, and 100) must be studied to evaluate whether the improved properties originate from graphene layers or the improvement of PMMA itself. According to the rule of mixtures (Equation 1 in Ref. 1), the changes in the evaluated strength values of graphene, $\Delta\sigma_{Gr}$, to alterations in the strength of PMMA stem from hot rolling, $\Delta\sigma_{PMMA}$, can be calculated as, $\Delta\sigma_{Gr} = -((1 - v_{Gr})/v_{Gr})\Delta\sigma_{PMMA}$. Note that the lower the volume fraction of graphene, $v_{Gr}$, the more sensitive it is to changes in PMMA strength resulting from hot rolling. Supplementary Fig. 1 depicts the sensitivity of the calculated strength for graphene in terms of the percentage enhancement of PMMA properties resulting from hot rolling for GPL with 10, 25, 50, 75, and 100 layers. The volume fraction of graphene is extracted from supplementary Table 1 of Ref. 1, and the strength of PMMA, without considering the possible additional strengthening from hot rolling, remains equal to the sole value provided for the strength of bare PMMA in Ref. 1, i.e., 79.6 MPa. As it can be seen, the reinforcing effect of hot rolling on PMMA results in a significant decrease in the evaluated strength of graphene and the decrease becomes larger with the reduction in number of graphene layers. In Ref. 1 the improvement in the strength of the GPL is attributed to the presence of graphene, the strength of graphene extracted by us from GPL with 10, 25, 50, 75, and 100 layers are determined as high as 23.18, 40.06, 32.23, 29.45, and 32.55 GPa, respectively (with the peak observed for 25 layers). *However*, significant effects of hot rolling on polymer strength have been reported in previous studies. For instance, in an experimental investigation on poly(oxymethylene) sheets (a thermoplastic polymer akin to PMMA in terms of strength), it was demonstrated that strength can be enhanced by 20% to 90% depending on process temperature and applied pressure[7]. Considering only a 20% improvement in PMMA properties due to hot rolling, the evaluated strength values for graphene is significantly decreased to exhibit as 5.47, 16.00, 18.26, and 24.18 GPa for 25, 50, 75, and 100 layers, respectively, while the 10 layer GPL shows a negative reinforcing effect (see the intersection of the vertical dashed line with the graphs in Supplementary Fig. 1). Thus, one can say that an improvement of only 23.1% in the strength of PMMA entirely removes any reinforcing effect of graphene in GPL-25.

Consequently, neglecting the influence of hot rolling on PMMA strength overestimates the derived values for the strength ($\sigma$) of graphene in Ref. 1. We suggest the enhancement in $\sigma$ of GPL likely happens entirely or almost entirely due to the improvement of $\sigma$ of PMMA through hot rolling.

Additional data presented in the "Peer Review File" (see Supplementary Fig. 2) of Ref. 1 shows GPL samples with constant PMMA thickness but varying numbers of graphene layers. It is worth noting that a moderate *decrease* in $\sigma$ (-4.6%) is observed for GPL-100 compared to GPL-50, despite having 100% increase in number of graphene layers, based on the data in Ref. 1. $E$ also *decreases* from GPL-10 to GPL-50 (-8%) and GPL-100 (-7.5%) according to Ref. 1. This suggests that the increase in $E$ and $\sigma$ reported in Ref. 1 for GPL-100 is due to thermal modification of the PMMA layer, not due to the presence of graphene.

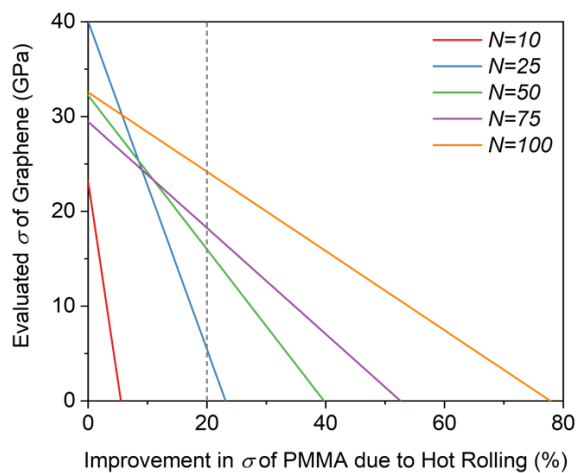

**Supplementary Figure 1:** *The range of estimated strength of graphene according to the percentage of improvement in the strength of PMMA due to hot rolling for the GPLs with 10, 25, 50, and 100 layers of graphene reported in Ref. 1.*

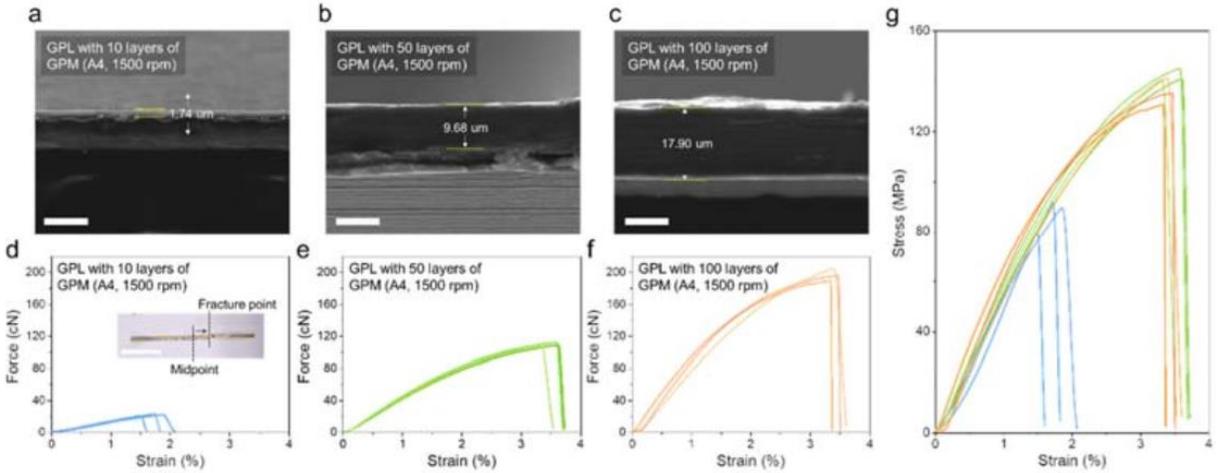

***Supplementary Figure 2:*** *Measurement of mechanical response of GPL with different layer number but with constant PMMA thickness, as presented in the "Peer Review File" of Ref. 1.*

**Section 2: Concept of Optimum Layer Number *N* in Layered Composites**

The concept of an optimal number of layers that maximizes a property is a well-known phenomenon in the study of layered composites[8]. This optimal point likely arises from the interplay between two competing interactions: Firstly, increasing *N* increases the total thickness (volume) of GPL, and consequently increases the number of defects, thereby decreasing strength. This decline in strength is a well-documented effect that can be mathematically described by the scaling law and the Weibull statistical equation. Conversely, increasing *N* and the thickness of GPL, modifies the temperature and pressure profile across the thickness of GPL. It affects the hot-pressing condition experienced by the graphene layer(s) and the graphene-PMMA interfaces across the thickness of GPL. Specifically, the graphene layers positioned near the two contact surfaces with hot rollers undergo extreme hot-pressing, while the inner layers experience a gradual reduction in applied hot-pressing. The competition between these two phenomena may justify the existence of an optimal $N$.

The concept of the optimal number of layers also emerges in the comparison of GPLs with varying volume fractions of graphene. For instance, GPL-25, with a volume fraction of 0.046%, exhibits the maximum strength for graphene, surpassing even GPL-100, which has a volume fraction four times greater. This calls for conducting further experiments and *meticulous* observations to elucidate the competing mechanisms that contribute to the optimal design of the GPL. The emergence of the optimal number of layers for GPL under hot rolling provides further evidence of the necessity to offer detailed observations of the effect of hot rolling on the properties of layered PMMA. Consequently, relying solely on bulk PMMA properties not only results in an inaccurate evaluation of graphene properties but also precludes the understanding of other strengthening mechanisms, such as the modification of interface strength across the thickness of GPL.

***Supplementary Table 1:*** *Comparison of thicknesses, as reported in Ref. 1 vs. our analysis.*

| SEM Micrograph | Thickness (reported in Ref. 1) | Thickness (our analysis) |
|---|---|---|
| Fig. 3c; S-GPL | Not mentioned | 3.77<br>3.61<br>3.60<br>Average 3.66 ± 0.10 µm |
| Fig. 3c; $T_g$-GPL | Not mentioned | 3.08<br>2.82<br>2.96<br>Average 2.95 ± 0.13 µm |
| Fig. 3c; GPL | 18.32 ± 0.63 µm<br>(as per supplementary Table 1 of ref.1) | 2.49<br>2.33<br>2.24<br>Average 2.35 ± 0.13 µm |
| Supplementary Fig. 5<br>GPL-0 | 18.64 ± 0.70 µm | 18.67<br>18.35<br>19.31<br>Average 18.78 ± 0.49 µm |
| Supplementary Fig. 5<br>GPL-10 | 18.35 ± 0.85 µm | 21.89<br>21.99<br>22.89<br>Average 22.26 ± 0.55 µm |
| Supplementary Fig. 5<br>GPL-25 | 18.61 ± 0.87 µm | 22.89<br>22.61<br>23.18<br>Average 22.89 ± 0.29 µm |
| Supplementary Fig. 5<br>GPL-50 | 18.72 ± 0.84 µm | 23.61<br>22.75<br>22.75<br>Average 23.03 ± 0.50 µm |
| Supplementary Fig. 5<br>GPL-75 | 18.95 ± 0.72 µm | 20.32<br>20.74<br>21.61<br>Average 20.89 ± 0.67 µm |
| Supplementary Fig. 5<br>GPL-100 | 18.32 ± 0.63 µm | 25.75<br>24.03<br>25.55<br>Average 25.11 ± 0.94 µm |
| Fig. 1c | 18.32 ± 0.63 µm<br>(as per supplementary Table 1 of ref.1) | 29.21<br>30.51<br>30.95<br>Average 28.42 ± 0.90 µm |

**Supplementary Table 2:** *Summary of calculated Young's Modulus (**E**) and Strength (**σ**) of graphene stacked PMMA layers.*

| Sample No. | S-GPL | Tg-GPL | GPL | GPL-0 |
|---|---|---|---|---|
| | \multicolumn{3}{c}{*E* (GPa)} | |
| #1 | 2.99 | 6.21 | 7.26 | 3.21 |
| #2 | 2.91 | 5.38 | 6.17 | |
| #3 | 2.59 | 5.67 | 6.34 | |
| **Average** | **2.83** | **5.75** | **6.59** | |
| | \multicolumn{3}{c}{*σ* (MPa)} | |
| #1 | 59.66 | 86.56 | 104.10 | 83.63 |
| #2 | 50.27 | 74.23 | 86.56 | |
| #3 | 46.82 | 87.22 | 111.99 | |
| **Average** | **52.25** | **82.67** | **100.88** | |